32ND INTERNATIONAL COSMIC RAY CONFERENCE, BEIJING 2011

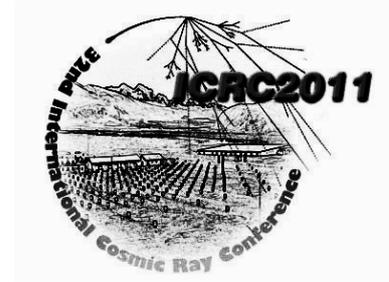

# $^{222}$Rn daughters influence on scaler mode of the ARGO-YBJ detector

ELIO GIROLETTI[1,2], IRENE BOLOGNINO[1,2], CLAUDIO CATTANEO[2], GIUSEPPE LIGUORI[2], PAOLA SALVINI[2], PIERO VALLANIA[3,4], CARLO VIGORITO[3,5]
[1] *Dipartimento di Fisica Nucleare e Teorica, Università degli Studi di Pavia, via Bassi 6, 27100 Pavia, Italy*
[2] *Istituto Nazionale di Fisica Nucleare, Sezione di Pavia, via Bassi 6, 27100 Pavia, Italy*
[3] *Istituto Nazionale di Fisica Nucleare, Sezione di Torino, via Giuria 1, 10125 Torino, Italy*
[4] *Istituto di Fisica dello Spazio Interplanetario dell'Istituto Nazionale di Astrofisica, corso Fiume 4, 10133 Torino, Italy*
[5] *Dipartimento di Fisica Generale, Università degli Studi di Torino, via Giuria 1, 10125 Torino, Italy*
elio.giroletti@pv.infn.it

**Abstract:** The ARGO-YBJ experiment is a full coverage air shower array; its lowest energy threshold is reached using the "scaler mode technique". Working in this mode, the signals generated by any particle hitting each cluster are put in coincidence every 150 ns and read by four independent scaler channels, giving the counting rates of multiplicity $\geq 1$, $\geq 2$, $\geq 3$ and $\geq 4$ (C1, C2, C3 and C4, respectively). The study of these counting rates pointed out a different behaviour of C1 respect to C2, C3 and C4, suggesting that C1 is detecting not only cosmic rays. This work shows that the radon ($^{222}$Rn) gamma emitter daughters present in the ARGO-YBJ building air are contributing to C1 counts at the level of 1 Hz each Bq/m$^3$ of radon. The uncertainty about this contribution is great, because of the high variability of $^{222}$Rn concentration and the building ventilation. The radon monitoring will allow the C1 correction improving the sensitivity of the ARGO-YBJ experiment at its lowest energy threshold.

## 1 Introduction

The Argo-YBJ experiment, located at Yangbajing (Tibet, P.R. China, altitude of 4300 m a.s.l.), is designed for VHE astronomy and cosmic ray observations at energies ranging from few hundreds Gev up to the knee region. It works in two modes: the shower mode and the scaler mode (see par. 2). In this last mode, the lowest multiplicity channel (C1 in the following) shows variations different from the ones of the higher multiplicity channels (C2, C3, and C4, in the following) and its average rate is higher than expected [2]. This behavior can be explained assuming C1 is detecting signals due not only to cosmic rays, as gamma emissions from natural radioactivity, both inside and under the ARGO-YBJ experimental hall.
In this work, we focus on the signal variability, studying this behavior with two different methods. Now we claim that gamma-emitter radon daughters contribute to C1 counts at a level of about 1% every 500 Bq/m$^3$ of $^{222}$Rn concentration in the experimental hall air.

## 2 The ARGO-YBJ detector

The detector consists of a single layer of RPCs operated in streamer mode grouped into modules called "clusters" (5.7x7.6 m$^2$). Each cluster is constituted by 12 RPCs. 130 clusters are installed to form the carpet of about 5600 m$^2$, with 93% of active area, surrounded by 23 additional clusters; the total instrumented area is of 100x110 m$^2$. Details about the detector and its performance can be found in [1]. The detector has two independent data acquisition systems, corresponding to the shower and scaler modes. In the scaler mode the total counts are measured every 0.5s: for each cluster signals are added up and put in coincidence in a narrow time interval (150 ns), giving the counting rates of $\geq 1$, $\geq 2$, $\geq 3$ and $\geq 4$, read by four independent scaler channels. These counting rates are referred in the following as C1, C2, C3, and C4, respectively. Their corresponding experimental average rates are ~40 kHz, ~2 kHz, ~300 Hz and ~120 Hz. The scaler mode technique allows the detection of transient emissions with an energy threshold of ~1GeV.
In a previous work [2], in order to correct the counting rates for atmospheric pressure and detector gas temperature variations, we found that the C1 channel performance is different from the ones of higher multiplicity channels (C2, C3 and C4). We are showing now that C1 time variations can be explained assuming that C1 is detecting local natural radioactivity, particularly the gammas emitted by radon daughters.

## 3 Radon gas in air

Radon, $^{222}$Rn, is a noble gas belonging from the uranium ($^{238}$U) radioactive family. It comes out from soil and enters buildings because its halflife is long enough (3.82 days). It produces radioactive isotopes that are gamma emitters: $^{214}$Bi, $^{214}$Pb, $^{214}$Pb and $^{210}$Pb. Radon concentration in indoor air is governed by many variables, such as: microclimatic conditions, building and soil characteristics [7,8,9]. Forecasting indoor radon concentration is very difficult inside a closed environment; predicting it at the ARGO-YBJ hall is impossible because of uncontrolled hall ventilation conditions. For these reasons, the radon concentration in the hall air is now continuously monitored both at the detector centre and near the north building wall (North side, in the following) using two Lucas cells (MIAM srl, Italy).

To quantify the radon influence on C1 counts, we made simulations with the Fluka code [5,6], about the detection of photons emitted by radon daughters in the hall, based on various $^{222}$Rn concentrations in air. The simulated RPC efficiency has been checked by means of radioactive sources, $^{60}$Co and $^{137}$Cs [3,4]. The expected contribution to C1 count rate due to γ-emitters radon daughters is about 1 Hz per Bq/m$^3$ of $^{222}$Rn concentration in air. However, even if these simulations show that radon gas can influence the scaler mode, they has to be considered as gross results, because of the high uncertainties related to the main parameters influencing C1 counts (hall ventilation, radon distribution inside the hall, radon daughter deposition over the detector, etc.). We expect that natural radioactivity influences only the C1 channel, being negligible the probability that ⩾2 γ's (emitted by radon daughters) are in coincidence within a time window of 150 ns.

## 4 Data analysis

Assessing radon gas contribution to C1, we used various methods to correlate our experimental time series x(t): radon gas in air (at North side and at detector centre), atmospheric pressure P, detector gas temperature T, and scaler counts C1, C2, C3 and C4.
In this work, we are using normalized time series $x(t)_{scld}$, defined as follows:
$x(t)_{scld} = [x(t) - <x(t)>]/\sigma(x(t))$
where x(t) is the data time series at time t, $<x(t)>$ and $\sigma(x(t))$ respectively are the average and the standard deviation, both calculated all over the examined period. Using normalized series, we can directly compare time series having different counting rates, such as C1, C2, C3 and C4 and radon data.
The analysis were performed with two methods: the linearization method and the proportional method.

Following the first method, C1 is assumed to depend linearly on the atmospheric pressure P and on the detector gas temperature T, according to the next equation:
$$C1(t) = a + b*P(t) + c*T(t) + C1_{RESIDUE}(t) \quad (1)$$
where the residual term $C1_{RESIDUE}$ is expected to be proportional to the radon concentration in air $C_{Rn}$. The correlation coefficient between $C_{Rn}$ and $C1_{RESIDUE}$ is expected to be high when the influence on C1 because of other physical phenomena than radon is negligible. Moreover, the proportion between the two series, $C1_{RESIDUE}$ and $C_{Rn}$, shows how much radon gas in air is affecting C1.

According to the second method (proportional method), radon influence can be evidenced subtracting to C1 two *a priori* "unknown" signals: the cosmic ray contribution γ1 and the background signal Bck. As first approximation, we can make the hypotheses that P and T are influencing the cosmic signal γ1, with the same proportion as in the other scalers (γ2, γ3 and γ4). We can write:
$$\gamma1(t) = k2*\gamma2(t) = k3*\gamma3(t) = k4*\gamma4(t) \quad (2)$$
where kn=<C1(t)/Cn(t)>, with n= 2, 3 and 4, and γ2=C2, γ3=C3 and γ4=C4 signals, in absence of other physical phenomena influencing the higher multiplicity channels. Thus, the radon contribution on C1, at time t, is given by $C1_{NET}$, according to the following equation:
$$C1_{NET}(t) = C1(t) - \gamma1(t) - Bck \quad (3)$$
where the Bck contribution is related to the detector background (which is mostly influenced by the soil natural radioactivity, as first approximation assumed constant) and γ1 can be calculated according to equation (2), starting from the higher multiplicity scalers. The relation between the two series, $C1_{NET}$ and $C_{Rn}$, represents how much radon gas is affecting C1.

## 5 Results and discussion

Radon gas concentration in the ARGO-YBJ hall air is highly variable in time, as expected. The average concentration at detector centre is 300–500 Bq/m$^3$, while at North side it reaches up to 5,000 Bq/m$^3$ in short periods[1]. For example, during the period from 7$^{th}$ to 19$^{th}$ July 2010, the radon concentration at the carpet centre was 450±194 Bq/m$^3$ (range 116–1058 Bq/m$^3$) while at North side it was 859±470 Bq/m$^3$ (range 330–2487 Bq/m$^3$). Radon behaviour is interesting: sometimes, it shows similar performance at both sites, other times there are great differences between them, because of fast increases of radon concentrations at North. In Fig.1, two radon time series (detector centre and North side) are reported (from 2$^{nd}$ to 15$^{th}$ June 2010), where fast radon concentration variations at North side are evident.
Given the surprisingly high concentrations at North side, it appears that radon is entering into the ARGO-YBJ building mainly from there and diffuses within the hall, where it is partially removed by ventilation. We are expecting that radon at centre position can be a better

---

[1] These concentrations appear high when they are compared to the maximum average concentration internationally admitted for indoor environments, 500 Bq/m$^3$

measurement of the average radon concentration over the whole ARGO-YBJ carpet.

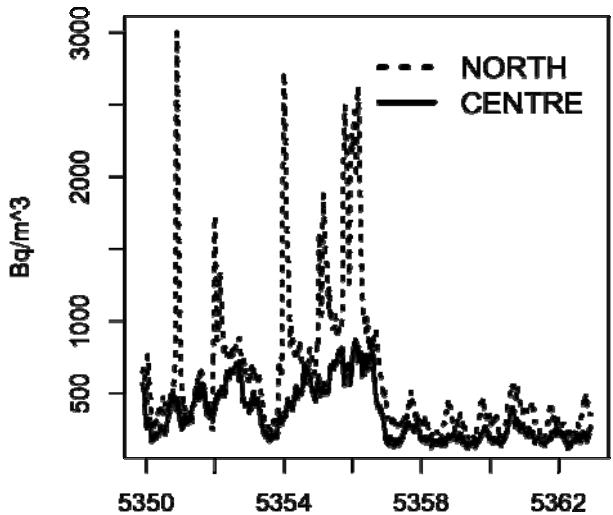

Figure 1. Radon gas concentration in air (Bq/m$^3$), measured at the detector centre (solid line) and at North side (dashed line). The time is in MJD, from 2$^{nd}$ to 15$^{th}$ June 2010; each measurement is every 0.5 hours.

We evaluated time series at different seasons of 2010 (from 1$^{st}$ January to 28$^{th}$ February; from 2$^{nd}$ to 15$^{th}$ June; from 1$^{st}$ October to 31$^{st}$ December), considering various clusters chosen depending on their position inside the experimental hall (at North side: 004, 032; in the middle: 104, 188, and at South side: 208 and 228).
In general, the correlation of C1$_{NET}$ and C1$_{RESIDUE}$ with the radon gas concentration at the detector centre is better for clusters at the carpet centre, and worse for the ones at the North and South sides. More again, the correlation of C1$_{NET}$ and C1$_{RESIDUE}$ with the radon concentration measured at North position is worst, being good only during periods when the North radon concentration doesn't have great variations.
For the same clusters, worst correlations are obtained during periods in which atmospheric high electric field variations are detected at the ARGO-YBJ site (see "Observation of the effect of atmospheric electric fields on the EAS with the ARGO-YBJ experiment", these Proceedings).

In the following the results of the linearization and proportional methods are described. Because radon data are collected every 0.5 hours, scaler counts, atmospheric pressure and gas temperature time series are averaged over the same interval.

*Method of linearization.* Fig.2, for the cluster 104, shows the similarity between C1$_{RESIDUE}$ and the radon concentration at the detector centre C$_{Rn}$, being the correlation coefficient between the two series 0.94. Moreover, the regression coefficient between C1$_{RESIDUE}$ and C$_{Rn}$ (representing the radon influence on C1) is 1.6 Hz/(Bq/m$^3$), in agreement with our simulations.
The method gives slightly different results when applied to other clusters: the lowest correlation coefficient between C1$_{RESIDUE}$ and C$_{Rn}$ has been found for a cluster belonging to the hall northern region. Table 1 shows the correlation coefficient between the radon concentration C$_{Rn}$ (at the detector centre) and C1$_{RESIDUE}$ and, in the third column, the regression coefficient between C1$_{REDISUE}$ and C$_{Rn}$, averaged all over mentioned periods.
The correlation coefficient doesn't improve using the radon North-side concentration, because its fast variations don't permit $^{222}$Rn to reach the secular equilibrium with γ–emitter daughters. Indeed, as previously mentioned, correlation with radon monitored at North side is in general worst than the one measured at the detector centre. As expected, all C2, C3 and C4 signals show bad correlation with radon concentration, both at centre and at North positions.

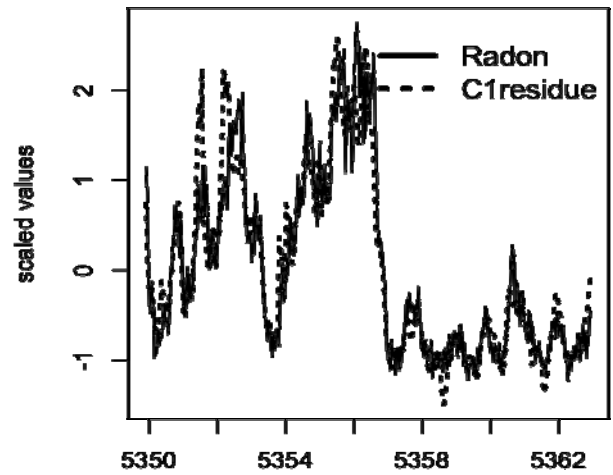

Figure 2. Normalised variations of the radon concentration (solid line) and C1$_{RESIDUE}$ (dashed line), for cluster 104. The time is in MJD, from 2$^{nd}$ to 15$^{th}$ June 2010; each measurement is every 0.5 hours.

| Cluster | correlation coefficient | C1$_{RESIDUE}$/C$_{Rn}$ Hz/(Bq/m$^3$) |
|---------|------------------------|---------------------------------------|
| 004     | 0.47                   | 0.79±0.04                             |
| 032     | 0.58                   | 1.02±0.04                             |
| 104     | 0.76                   | 1.29±0.03                             |
| 188     | 0.77                   | 1.18±0.02                             |
| 208     | 0.57                   | 0.78±0.03                             |
| 228     | 0.42                   | 0.71±0.04                             |

Table 1. Results of the method of linearization (see text).

*Proportional method.* In Fig.3, for the cluster 104, the normalized radon concentration at the centre position is compared with C1$_{NET}$ (C1 proportionally corrected with C2 counts, according to equation 3, k2=9.8±0.4). For the data of Fig.3, the maximum correlation coefficient between C$_{Rn}$ and C1$_{NET}$ is 0.92 once the Bck contribution of 24 kHz is subtracted.
Similar results are obtained by correcting C1 signal with the different multiplicity scalers, C2, C3 and C4. Concerning the regression coefficient between C1$_{NET}$ and C$_{Rn}$ at the detector centre, we obtained 1.65, 1.68 and 1.67 Hz/(Bq/m$^3$), when C1 is proportionally corrected with C2, C3 and C4, respectively.
In Table 2, the results obtained with this method are reported, averaged all over the analyzed periods. The Table shows the correlation coefficient between the radon concentration $C_{Rn}$ (at the detector centre) and $C1_{NET}$ and, in the third column, the regression coefficient between $C1_{NET}$ and $C_{Rn}$, averaged all over mentioned periods. The results are in agreement with the ones obtained with the linearization method.

The interesting aspect of this method is that the analysis is evidencing a contribution to C1 from the soil natural radioactivity of the order of 20±5 kHz (depending on different periods and clusters), in agreement with previous results [2]. A dedicated study is currently in progress.

ured at North position is generally lower, being high only during periods when the radon gas concentration doesn't have great variations.

To complete the analysis we applied the same methods to the higher multiplicity channels. As expected, we don't find any radon influence on the C2, C3 and C4 counting's.

Radon gas concentration in air is now being continuously monitored inside the experimental hall, in order to be able to correct the ARGO-YBJ detector counts for the radon influence, improving the sensitivity at its lowest energy threshold.

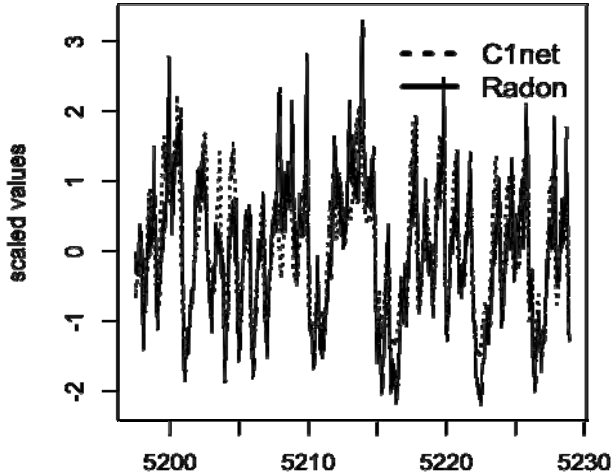

Figure 3. Normalised variations of radon concentration (solid line) and $C1_{NET}$ (dashed line) calculated using C2 (see text), for cluster 104. The time is in MJD, from 1st January to 1st February 2010; each measurement is every 0.5 hours.

| Cluster | Correlation coefficient | $C1_{NET}/C_{Rn}$ Hz/(Bq/m$^3$) |
|---------|------------------------|-------------------------------|
| 004 | 0.83 | 1.45±0.03 |
| 032 | 0.89 | 1.43±0.03 |
| 104 | 0.79 | 1.62±0.05 |
| 188 | 0.78 | 1.01±0.03 |
| 208 | 0.50 | 0.63±0.04 |
| 228 | 0.54 | 0.67±0.03 |

Table 2. Results of the proportional method (see text).

## 6 Conclusions

Natural radioactivity in air influences the ARGO-YBJ single counting rates at the level of about 0.5-1.7 Hz per Bq/m$^3$ of $^{222}$Rn concentration, because of the γ's emitted by its daughters. The average radon influence is about 1-3% of C1 counting's, as expected by Fluka simulations.

Some differences between clusters and periods are consistent with the strong variability, either in time as well as in space, of the radon gas and of its daughter concentrations in the air of the ARGO-YBJ hall.

In general, the correlation with radon concentration measured at the detector centre is higher for clusters installed at the carpet centre than for the North and South clusters. The correlation with radon concentration meas-